\documentclass[]{spie}  

\usepackage{amsmath,amsfonts,amssymb}
\usepackage{graphicx}
\usepackage[colorlinks=true, allcolors=blue]{hyperref}

\title{The SOXS scheduler for remote operation at LaSilla: Concept and design}

\author[a,b]{Marco Landoni}
\author[c]{Dave Young}
\author[d]{Laurent Marty}
\author[a]{Laura Asquini}
\author[c]{Stephen J. Smartt}
\author[a-e]{Alberto Trombetta}
\author[a]{Sergio~Campana}
\author[f]{Riccardo~Claudi}
\author[d]{Pietro~Schipani}
\author[a]{Matteo~Aliverti}
\author[f]{Andrea~Baruffolo}
\author[g]{Sagi~Ben-Ami}
\author[f, h]{Federico~Biondi}
\author[d]{Giulio~Capasso}
\author[k]{Rosario~Cosentino}
\author[i]{Francesco~D'Alessio}
\author[a]{Paolo~D'Avanzo}
\author[l]{Ofir	Hershko}
\author[m, n]{Hanindyo~Kuncarayakti}
\author[o]{Matteo~Munari}
\author[p, q]{Giuliano~Pignata}
\author[h]{Adam~Rubin}
\author[o]{Salvatore~Scuderi}
\author[g]{Fabrizio~Vitali}
\author[l]{Jani~Achrén}
\author[p,q]{José~Antonio~Araiza-Duran}
\author[n]{Iair~Arcavi}
\author[r]{Anna~Brucalassi}
\author[h]{Rachel~Bruch}
\author[f]{Enrico~Cappellaro}
\author[d]{Mirko~Colapietro}
\author[d]{Massimo~Della~Valle}
\author[f]{Marco~De~Pascale}
\author[o]{Rosario~Di~Benedetto}
\author[d]{Sergio~D'Orsi}
\author[l]{Avishay~Gal-Yam}
\author[a]{Matteo~Genoni}
\author[k]{Marcos~Hernandez}
\author[m, n]{Jari~Kotilainen}
\author[s]{Gianluca~Li~Causi}
\author[n]{Seppo~Mattila}
\author[l]{Michael~Rappaport}
\author[f]{Kalyan~Radhakrishnan}
\author[f]{Davide~Ricci}
\author[a]{Marco~Riva}
\author[f]{Bernardo~Salasnich}
\author[o]{Ricardo~Zanmar~Sanchez}
\author[t]{Maximilian~Stritzinger}
\author[k]{Hector~Ventura}

\affil[a]{INAF -- Osservatorio Astronomico di Brera, Via Bianchi 46, I-23807, Merate, Italy}
\affil[b]{INAF - Istituto Nazionale di Astrofisica. Osservatorio Astronomico di Cagliari, Via della Scienza 5, I-09047 Selargius (CA), Italy }
\affil[c]{Astrophysics Research Centre, School of Mathematics and Physics, Queen's University Belfast, Belfast BT7 1NN, UK}
\affil[d]{INAF -- Osservatorio Astronomico di Capodimonte, Salita Moiariello 16, I-80131, Naples, Italy}
\affil[e]{Dipartimento di Scienze Teoriche e Applicate, Universit\`a degli Studi dell’Insubria, Via Ottorino Rossi, I-21100 Varese, Italy}
\affil[f]{INAF -- Osservatorio Astronomico di Padova, Vicolo dell’Osservatorio 5, I-35122, Padua, Italy }
\affil[g]{Harvard-Smithsonian Center for Astrophysics, Cambridge, USA }
\affil[h]{Max-Planck-Institut f\"ur Extraterrestrische Physik, Giessenbachstrasse 1, D-85748 Garching, Germany }
\affil[k]{FGG-INAF, TNG, Rambla J.A. Fernández Pérez 7, E-38712 Bre\~na Baja (TF), Spain }
\affil[i]{INAF -- Osservatorio Astronomico di Roma, Via Frascati 33, I-00078 M. Porzio Catone, Italy }
\affil[l]{Weizmann Institute of Science, Herzl St 234, Rehovot, 7610001, Israel }
\affil[m]{Finnish Centre for Astronomy with ESO (FINCA), FI-20014 University of Turku, Finland}
\affil[n]{Tuorla Observatory, Dept. of Physics and Astronomy, FI-20014 University of Turku, Finland }
\affil[o]{INAF -- Osservatorio Astrofisico di Catania, Via S. Sofia 78 30, I-95123 Catania, Italy }

\affil[p]{Universidad Andres Bello, Avda. Republica 252, Santiago, Chile }
\affil[q]{Incident Angle Oy, Capsiankatu 4 A 29, FI-20320 Turku, Finland }

\affil[r]{ESO, Karl Schwarzschild Strasse 2, D-85748, Garching bei München, Germany}
\affil[s]{INAF - Istituto di Astrofisica e Planetologia Spaziali, Via del Fosso del Cavaliere 100, I-00133 Roma, Italy}
\affil[t]{Dark Cosmology Centre, Juliane Maries Vej 30, DK-2100 Copenhagen, Denmark }
\affil[n]{Tel Aviv University, Department of Astrophysics, 69978 Tel Aviv, Israel }

\affil[p]{Aboa Space Research Oy, Tierankatu 4B, FI-20520 Turku, Finland}

\affil[t]{Millennium Institute of Astrophysics (MAS)}
\affil[u]{Aarhus University, Ny Munkegade 120, D-8000 Aarhus, Denmark }

\authorinfo{Further author information: Send correspondence to M. Landoni\\ E-mail: marco.landoni@inaf.it}

\begin{document} 
\maketitle

\begin{abstract}
In this paper we present the SOXS Scheduler, a web-based application aimed at optimising remote observations at the NTT-ESO in the context of scientific topics of both the SOXS Consortium and regular ESO proposals.
This paper will give details of how detected transients from various surveys are inserted, prioritised, and selected for observations with SOXS at the NTT while keeping the correct sharing between GTO time (for the SOXS Consortium) and the regularly approved observing time from ESO proposals. For the 5-years of operation of SOXS this vital piece of software will provide a night-by-night dynamical schedule, allowing the user to face rapid changes during the operations that might come from varying weather conditions or frequent target of opportunity (ToO) observations that require a rapid response. 
The scheduler is developed with high available and scalable architecture in mind and it implements the state-of-the-art technologies for API Restful application like Docker Containers, API Gateway, and Python-based Flask frameworks.

\end{abstract}


\section{INTRODUCTION}
\label{sec:intro}  

The Son Of X-Shooter (SOXS) instrument\cite{soxs1,soxs2} is a single object spectrograph with spectral resolution R$\sim$ 4,500 to be mounted at the European Southern Observatory New Technology Telescope (ESO-NTT) in La Silla and operated in the ESO La Silla-Paranal Observatory operation environment without an astronomer on the mountain.
For this reason, it will be absolutely necessary to design and develop software able to aid the operator on the mountain 
to:
\begin{itemize}

\item{properly organize the night in advance;}
\item{prepare on-the-fly a backup schedule for the night by convolving the desired (and observable) targets with the current weather and pointing restriction conditions;}
\item{define a schedule that combines both Consortia Targets Target of Opportunity (ToOs, coming from the Marshall application, see below, in terms of classification targets and follow-up targets) and fast response events, and regular ESO proposals (submitted with P2, possibly including ToOs);}
\item{keep track of the operations of the instrument each night in terms of duty cycle, percentage of completed proposals and hit/miss rate of transient events coming from the Consortium targets.
}
\end{itemize}
The scheduler must communicate automatically with the official ESO Phase 2 tools to define Observing Blocks (OBs) from the targets in the standard ESO way and populate the Visitor Execution Sequence (VES) according to the visibility of the target, their priority and current weather condition (accessed through the Astronomical Site Monitor - ASM - Weather Application Programming Interface - API). A sketch of the Scheduler Application, required for remotely operate SOXS smoothly, is reported in Figure 1.

\begin{figure}
\begin{center}
\includegraphics[width=0.95\textwidth]{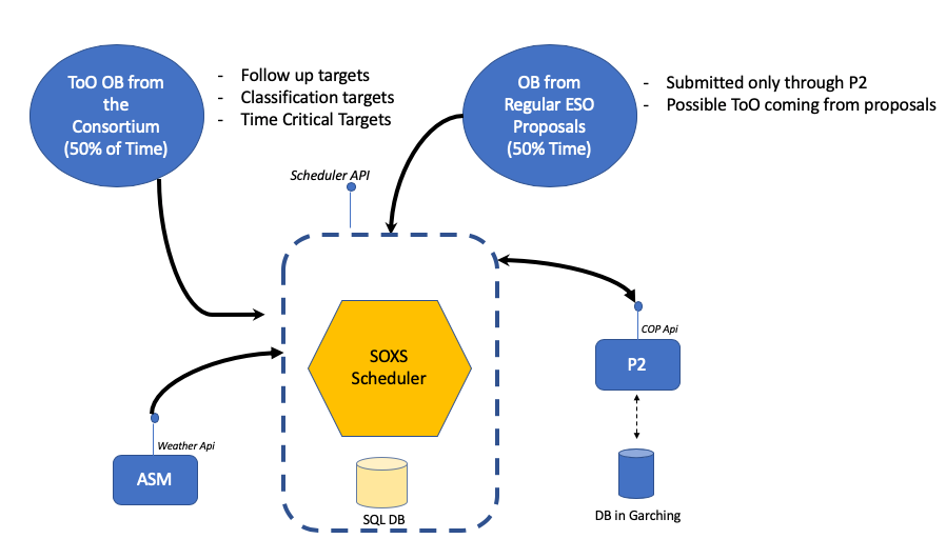}
\caption{Sketch concept for the SOXS scheduler.}
\label{flowch}
\end{center}
\end{figure}

The Scheduler will communicate and fetch OBs from two actors:
\begin{itemize}
    \item{The Marshall, which is a web-based application that collects and aggregates data from various transient surveys. OBs generated through the Marshall will be called Classification Targets OB (CTOB) or Follow-up OB (FUOB), according to the logic reported in 3.2.}
  \item{The regular ESO proposals, filled by Principal Investigators on the ESO P2 web page.}
\end{itemize}

The Scheduler will maintain the local cache and stateful representations of all the OBs that could be observed by SOXS during each semester, and acts as middle-ware with the ESO P2.
\section{The Marshall Application - Brief review}

The purpose of the Marshall Web Application is to collect data from various surveys around the world and produce a `ticket' for each unique transient source.
Each ticket in the Marshall has a unique numeric ID (giving the scheduler a reliable identifier for the detected transient) and is responsible for aggregating all data coming from various surveys in a single form (magnitude, stamps, comments or classification reports, etc.) and keeping track of the source.
We will adapt the already available Marshall for the ESO extended Public ESO Spectroscopic Survey of Transient Objects (ePESSTO \cite{epessto}) survey in the following ways:
\begin{itemize}
\item{define a new data flow for sources that enter the Marshall that match the requirements of SOXS;}
\item{define proper interfaces and an API to communicate with the SOXS scheduler, automating as much as possible the manual activities of the ePESSTO Target Alert Team (TAT).}
\end{itemize}

\begin{figure}
\begin{center}
\includegraphics[width=0.95\textwidth]{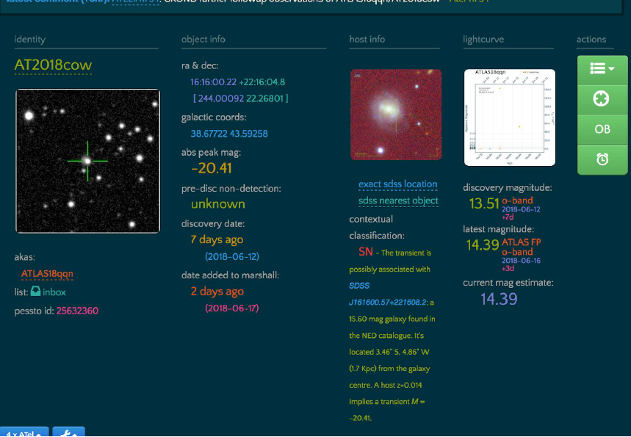}
\caption{An example of a Ticket in the ePESSTO Marshall (see e.g. Young et al. 2014\cite{epessto}).}
\label{fig:marshall}
\end{center}
\end{figure}

In Figure \ref{fig:marshall} we show a screenshot of a Ticket present in the ePESSTO Marshall, which displays basic information about the transient AT2018cow.
By automatically aggregating data from ATLAS, TNS and SDSS, the Mashall has produced a single ticket displaying all information needed by the TAT and producing the OBX files for observations (with the old P2PP) used by the NTT EFOSC2 instrument at La Silla. For the purpose of this paper it is not necessary to deeply introduce the logic behind the Marshall. Nevertheless, it is important to fix the main points about the flow of the sources in this application (which is external with respect to the scheduler), since they define precisely the interfaces with the scheduler for creating, updating and deleting OBs from the scheduler.

\subsection{The Workflow of sources in the Marshall}

\begin{figure}
\begin{center}
\includegraphics[width=0.95\textwidth]{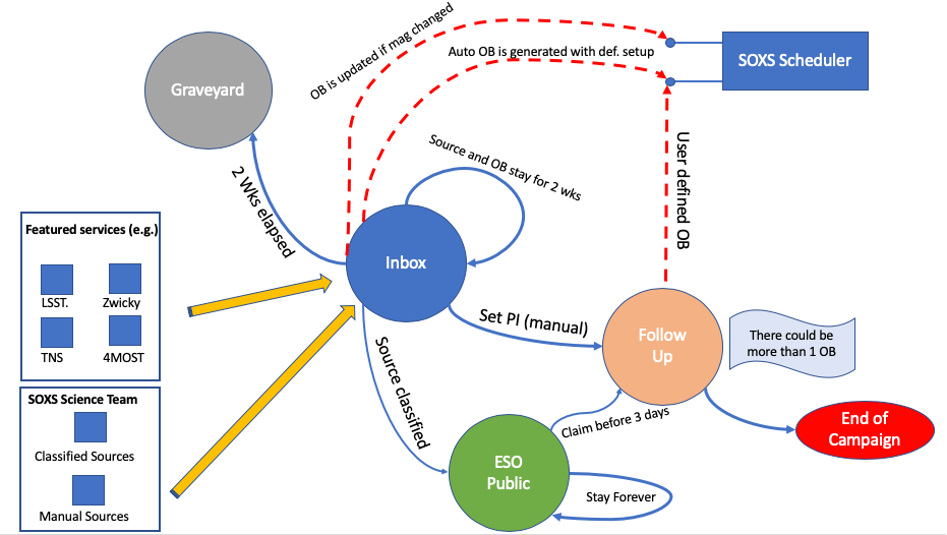}
\caption{Source data-flow in the SOXS Marshall: blue arrows model the flow of the targets between states, while dashed red arrows indicate the data-flow within the SOXS Scheduler. Data from surveys or from links external to the SOXS Marshall are drawn with solid yellow arrows.}
\label{fig:workflow}
\end{center}
\end{figure}

Figure \ref{fig:workflow} displays the flow of sources through the SOXS Marshall system. As explained in Section 3.2 below, the Marshall is intended to be used by the Consortium in order to:
\begin{itemize}
\item{keep track and continuously aggregate data from transient surveys (e.g. LSST, ATLAS, etc.) of newly discovered objects (or, in general, transient events);}
\item{define a good set of targets, for which the classification is still unknown, to be followed up spectroscopically with SOXS (referred as `classification targets');}
\item{mark some important targets for a dedicated follow-up spectroscopic campaign with SOXS (follow-up targets);}
\item{manually insert very specific high-priority objects (e.g. gravitational wave events) to trigger and urgent follow-up observations with SOXS in La Silla (manual targets).}
\end{itemize}

Tickets are created starting from the Feeder Surveys (FS). Feeder Surveys (e.g. Zwicky, LSST, etc.) export their data in some machine readable format at regular time intervals. The data contains information about new transients (RA, DEC, mag, etc.) or updates of already discovered transients (e.g. if the magnitude changes). When a new transient is discovered, a ticket is created with a unique Marshall Identifier (MID) and put in the Inbox state. If the transient is brighter than 19.50 mag and visible from La Silla, an OB is automatically sent to the SOXS Scheduler (which then interfaces with the ESO P2).
This OB is sent using the API createAutoOB (see Scheduler API section), which basically tells the scheduler that a new transient has been discovered and queued for observations as a Classification Target. From here, it stays in the Inbox state and in the queue of the Scheduler for 3 weeks. After this time window, if there was no possibility to observe the source, the OB is automatically removed from the scheduler (through the API autoCleanOBs) and the source is moved to the Graveyard state. Otherwise, if the source is observed during any night of the three weeks it is put in the ESO Public state and, if no one claims it, the source stays there forever.

If a source in Inbox appears to be particularly interesting, a person in the Consortium can claim the PI-Ship putting it in the follow-up state. From here new OBs (more then one if needed) are created via the API createFollowUpOB and sent to the scheduler. It's important to note that, when a source is in the follow-up state, the SOXS Marshall Webapp allows the PI to produce OBs with fine-grained setup (slit width, integration time, etc.) while, in the Inbox state, a standard setup is used. When the follow-up campaign of the source is completed, the PI moves the source in the Marshall to the end of campaign state. 
Finally, a source may be moved to the follow-up state from the ESO public state (having first been spectroscopically classified) if a person in the Consortium claims the rights on this source as PI, within 3 days from the classification taking place (if the classified source falls within the triggering criteria of the SOXS Consortium program).
It is also important to note that the magnitude of transients may change during time when an OB has already be issued to the scheduler. In this case, the magnitude and the exposure time of the OB are updated using the scheduler API updateAutoOBMagnitude. The the unique link between the OB and the source is assured by the uniqueness of the MID number. Moreover, if a source is as classified by some survey other than SOXS (e.g. 4MOST) the OB of classification target is deleted using the API deleteOB.  

Concerning very time critical observations (e.g. a gravitational wave trigger follow-up or a particular gamma-ray burst not caught by other surveys) a user may manually add a new transient ticket from the Marshall. After submission the transient appears in the follow up state and is fetchable by the scheduler.  

\section{P2 and weather API interfaces}

The SOXS scheduler will interface with the ESO-P2 in order to populate the Visitor Execution Sequence (VES). This is a sorted list of OBs that can be fetched directly with the Broker for Observation Blocks (BOB) by the operator and is fully integrated in the Control Software Environment provided by ESO.
The VES is the main concept of the night plan when observations are carried out at the telescope in Visitor Mode or Remote Visitor Mode. These observing modes are the only available setups for observation at La Silla, at the moment of the writing of this paper.

\begin{figure}
\begin{center}
\includegraphics[width=0.95\textwidth]{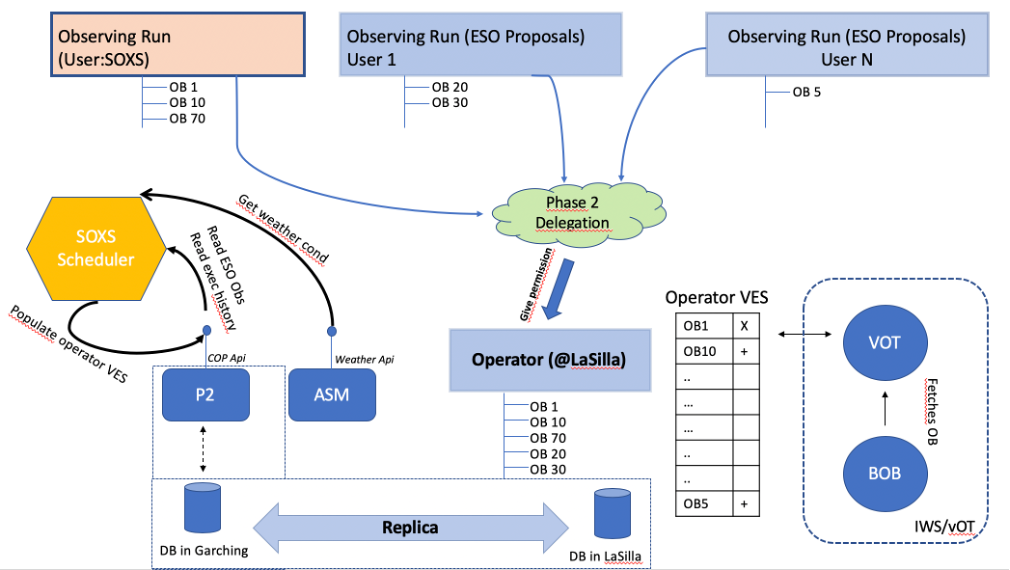}
\caption{SOXS Scheduler interface with the P2 and Weather API.}
\label{fig:p2}
\end{center}
\end{figure}

As reported in Figure \ref{fig:p2} the SOXS scheduler (in yellow) must interact with different application and concepts stemming from the new ESO P2 webapp (via API) and the weather API, which basically reports the current atmospheric conditions (such as seeing, wind direction, etc.) for observations.
Allowing the scheduler to dynamically populate the VES before the night and update it when observational conditions change, requires us to define a particular User (to deal with the SOXS Marshall Consortium target) and understand the concept of Phase 2 Delegation. Here it is reported how the schema of Figure 4 permits to solve the problem.
In the ESO P2, the following users shall be defined:
\begin{itemize}
\item{User Soxs: This user is the owner of the OBs that are uploaded to the ESO P2 (via API) though the Marshall.}
\item{For each PI of regular proposal accepted by ESO, there will be a user which owns its proper OBs uploaded to the ESO P2 via the Web interface of API.}
\item{User Operator: This user models the operator on the mountain and must be able to access the OBs from both the User SOXS and all the other users of regular ESO proposals.}
\end{itemize}

The Scheduler will only populate the VES of the Operator user.
To achieve this, the Soxs and other users created starting from regular accepted proposals must allow the Operator User to access their own OBs. This requires that both Soxs user (by default) and the various ESO-PI use the Phase 2 Delegation feature. Briefly, this feature offered by the P2 allows a user to gain access to the OBs from another PI in the P2 system. 
For example, as seen in Figure \ref{fig:p2}, the Operator user can see OB1, OB10 and OB70 from Soxs user as well as OB20 and OB5 from User 2 and User N. This is crucial, for the Scheduler, to populate the VES night by night and update it according to the current weather conditions.
As a baseline, in the European afternoon the Scheduler provides an initial schedule computed following the strategy reported in the next section, taking into account the scientific prioritization of the targets and the static fulfillment of the constraints (airmass, visibility, fraction of lunar illumination and Moon distance). Throughout a Web page developed ad-hoc, the SOXS science team validates the proposed schedule and the OBs that compose it get synchronized between the local scheduler database and the P2 current status, and pushed into the operator VES. 
The VES is checked by the operator regularly against current weather conditions, weather changes and possibly pointing restrictions. If one or more OBs in the current VES cannot be observed anymore, the scheduler selects backup targets from the visible ones, convolving the new constraints with the current weather measured by the ASM API, and the VES is subsequently updated. This procedure shall require no more than 1 minute in order to reduce as much as possible the loss of time at the telescope. 

\section{The scheduling algorithm}
\subsection{General description}
The scheduling algorithm is developed in Python 3 and uses tools from Astropy libraries\cite{robitaille2013astropy, price2018astropy}, such as astroplan\cite{morris2018astroplan}. Astroplan is a tool designed for observation planning, and it is provided with common observational quantities such as celestial rise and set times at a given Earth location. For the SOXS scheduler, this tool is used to compute observational constraints for the targets, and one of its built-in schedulers is employed. 
The necessity for a fast response and updates, given both the nature of the targets, and for the numerous interfaces the scheduler has with external APIs and applications, made us divide the algorithm's workflow into two steps (Fig. \ref{flowch}). The first step consists in a selection of those targets that potentially can be observed during the designated night, and it handles most of the computational workload. As such, this step is performed only once per night and requires no update on the run. The second step is the actual scheduling process, that will be examined in section \ref{sched}. 

\begin{figure}[ht]
\begin{center}
\includegraphics[width=0.55\textwidth]{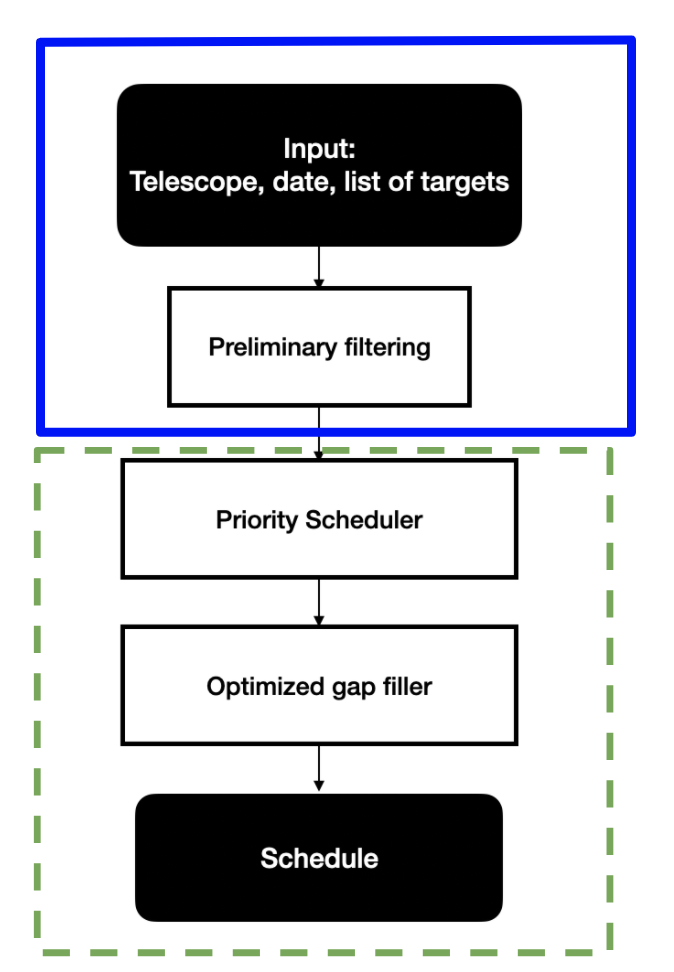}
\caption{Sketch of the scheduler workflow: the blue solid box represents the operations that will be executed once per night; the green dashed box highlights the operations that can be executed multiple times.}
\label{flowch}
\end{center}
\end{figure}

\subsubsection{Filtering of targets}
A preliminary filtering of targets is needed to select a pool of objects that are in principle observable during the designated night. With the term “observable” we mean the time range during the night in which all of the visibility constraint of the single object are satisfied, and that this time range is at least as long as the required exposure. Since the filtering is performed once before each night, it is convenient to make it as detailed as possible in order to reduce the workload on the actual scheduler. The algorithm fetches the targets from the database which contains all the information about the objects, such as name, magnitude, coordinates and so on. Then, it evaluates the constraints imposed by the user. In order to do so, it relies on the applicable built-in astroplan classes (e.g. AirmassConstraint(), AtNightConstraint(), MoonSeparationConstraint()). Except for the constraints which are independent of the targets (such as the fractional lunar illumination), the other classes have a method that takes as an input an array of timestamps at which to compute the constraints, the location on Earth of the observer (in our case the NTT) and the targets coordinates. These methods return a 2D array for each constraint, with the score of the evaluation on one axis and the correspondent timestamp on the other. Once all the constraints are computed, they are collected in a constraint list for future use during scheduling. The algorithm then turns each 2D constraint array into a 1D boolean array, where each slot represents a timestamp (with a spacing of 5 minutes between each) and its value is True if the constraint is satisfied at that moment, False otherwise. Then, the constraints are convolved together to return a single 1D array of zeroes and ones with the same time spacing, and whose values are 0 if at least one constraint is not satisfied at that moment and 1 only if all of them are. This final array is called the observability array and it will be fundamental for scheduling purposes. The final test to check if the object can be actually observed is to compare the length of the time span at which the object is observable (the observability array presents a 1) and the exposure required for observations. In order to do so, the number of 5-minute slots (nslots hereafter) that the exposure would require is computed (taking the ceiling integer value) and the observability array is checked to show at least nslots of contiguous slots containing a 1. If the filtering process is successfully passed by a target, the target is associated with a score that represents an overall evaluation of the level of satisfaction of the constraints during that night. This score is evaluated by taking into account various cumulative probability functions already used in survey scheduling\cite{bierwirth2010new}. After this step, the object is stored as a dictionary in a tonight bucket list, that contains: 

$\bullet$ name of the target;

$\bullet$ coordinates (RA, Dec);

$\bullet$ observability array;

$\bullet$ nslots;

$\bullet$ exposure;

$\bullet$ priority;

$\bullet$ constraint list;

$\bullet$ score.

The tonight bucket is then sorted by score and passed to the actual scheduler.

\subsubsection{Scheduling}
\label{sched}
In order to deal with the scientific prioritization of the targets, which is especially important for follow-up targets, and ESO proposals, the built-in astroplan priority scheduler is implemented as a first step and deals with the top priority targets. This scheduler is designed to find the optimal (given the constraints) observing time for each object in order of priority. This algorithm sorts the observing blocks by priority and evaluates the constraints for each one with the very fine time resolution of one minute. Then, for the single object under examination, it assigns a numeric score to each timestamp of each constraint. The score is a number between 0 and 1 representing the level of satisfaction of the observing conditions at that moment, relatively to that specific constraint. The product of the scores associated to the same timestamp is then calculated and the time relative to the highest score is whenever the algorithm attempts to schedule the object. If that spot is already occupied, the second best spot is tried, and so on. This process is very efficient in optimizing the observing conditions, but it is computationally quite heavy because of the time resolution at which the constraints are evaluated, hence the decision to employ this scheduler only for top priority objects. The task now remains to optimally make use of the remaining time during the night. The algorithm is built around a parametrization of the night that allows a direct comparison between the observability array of each target and the gaps available for scheduling. By using the date of the desired night and the coordinates of the observing site, the exact times of the astronomical twilight (evening and morning) can be retrieved, hence the duration of the night. Then, the night is parametrized by a night array, whose length is the ceiling integer of this time span divided by the chosen time resolution (5 minutes). This way, the length of the night array and that of the observability arrays are the same, each slot representing the same time span during the night. The values inside each slot in the night array can be either 0, if the slot is not occupied by an observation, or 1 otherwise, so that at the beginning of the night this array shows zeroes only.
The schedule that the priority scheduler returns as a list of ObservingBlocks and TransitionBlocks is converted into a list of Python dictionaries with all of its previous information (as in the tonight bucket), plus a new value in the dictionary, containing information about the filled slots. Such information are the indices of the slots that in the night array are changed from 0 to 1, as they are occupied by these observations. From this point on, all future processes will deal with a night array that is partially free and a list of targets (proto-schedule) to complete. This fixed parametrization for the night is indeed more rigid than the one employed by astroplan schedulers, since blocks cannot be created on the fly, but with a proper choice of time resolution it can balance a lighter computational workload (i.e. faster response) with a good estimate of observational constraints. 

 \begin{figure} [ht]
\centering
\includegraphics[width=0.8\textwidth]{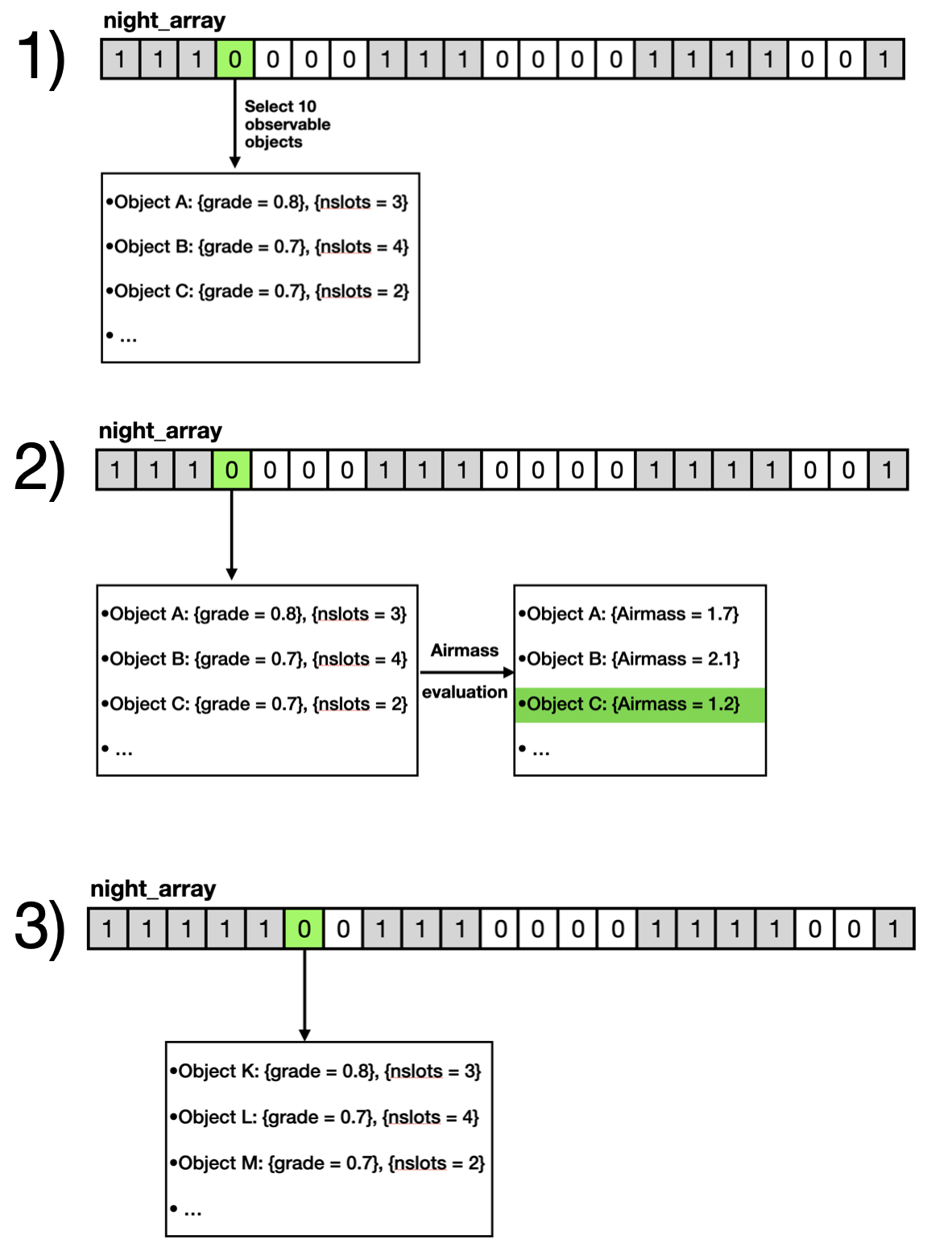}
\caption{Visualization of the workflow of the gap filler function.}
\label{gggf}
\end{figure}

To fill the gaps left by the priority scheduler (Fig. \ref{gggf}), the algorithm looks for the first free slot(s) in the night array and finds in the tonight bucket the first objects that fit the gap, by checking their observability array and slots (we fix the number of fitting objects equal to 10). Having found such objects, the algorithm evaluates the constraints at the mean time of the candidate observation and chooses the best one. The winner of this competition is appended to the schedule and deleted from the tonight bucket, and the night array is changed accordingly. This process is repeated until the night is full, or there are no objects in the tonight bucket that can fill the gaps. At the end, the schedule is sorted by slots occupied. 

\subsection{On the fly update}
\begin{figure}[!ht]
\centering
\includegraphics[width=0.8\textwidth]{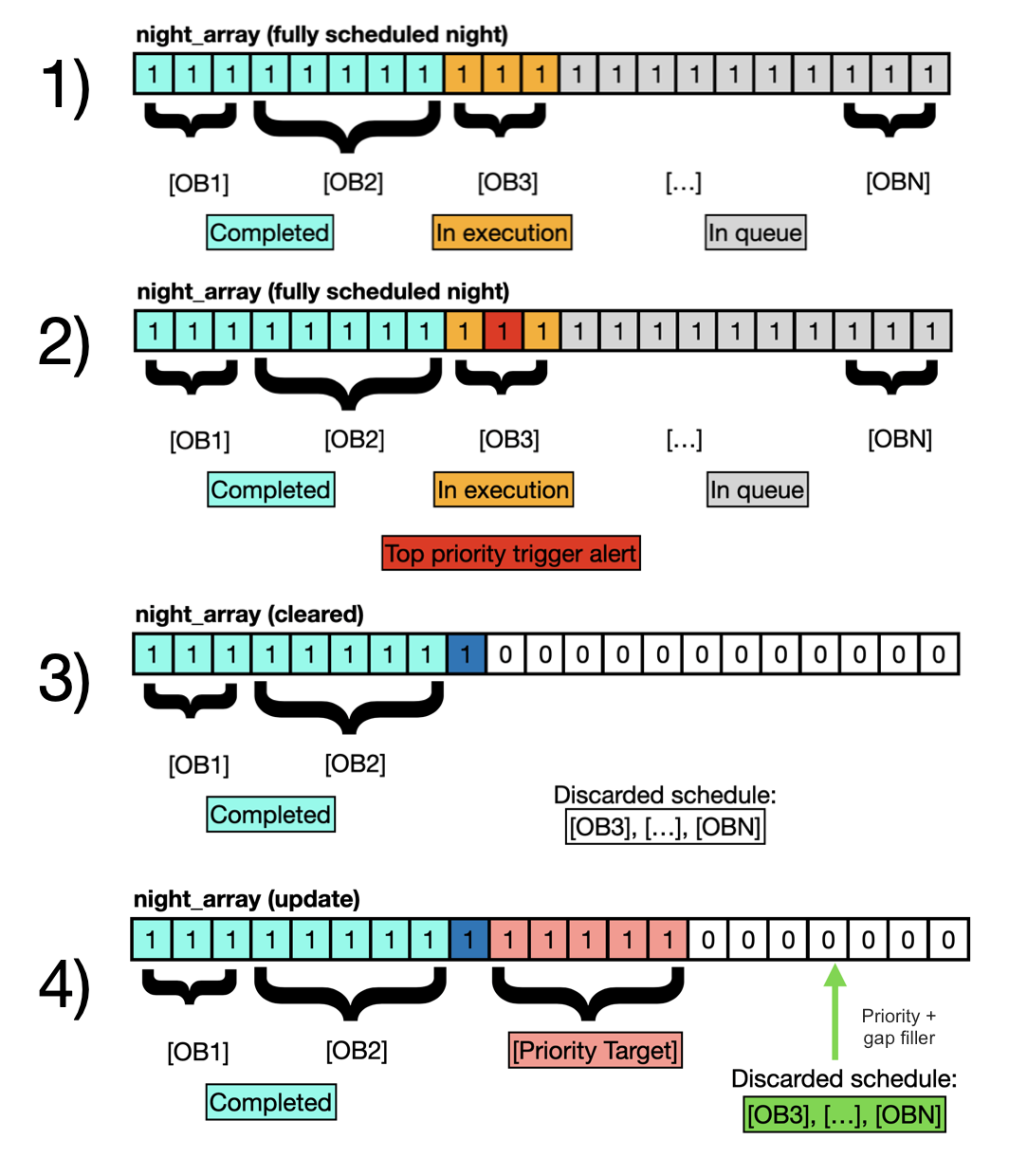}
\caption{Visualization of the workflow for updating of the schedule, from the arrival of the trigger to the final rearranging of the night.}
\label{forceob}
\end{figure}

The scheduling process described above can be applied also to deal with on the fly updates, that can be triggered by changes in weather conditions or by the arrival of a top priority target. In case of updates that involve the sudden impossibility of observing certain targets, the corresponding slots in the night array can be cleared, the targets deleted from the schedule and the scheduling algorithm run again on the partially empty schedule. Similarly, in case of arrival of a priority trigger (Fig. \ref{forceob}), the idea is based on the optimization process that can update the schedule with the new event. At the arrival of the trigger, the current time and slot in the night array is registered. From that slot the night array is cleared and the top priority event occupies the first free slots (as many as it requires). The time corresponding to the end of its observation is registered and all of the unobserved objects in the original schedule (including the possible target whose observation was interrupted by the trigger), are passed again to the priority scheduler and the gap-filler function. A further check for gaps in the schedule is performed and, if any is found, it is filled with the gap-filler function, ensuring that no time during the night is wasted, while keeping the good quality of the pre-trigger schedule.

\section{API and Web application}
\label{sec:webapp}
The SOXS Scheduler will expose its services through RESTful API and it will offer two main features:

$\bullet$ {a restful API based system;}

$\bullet$ {a web application.}


For the former, we designed a RESTful API for the scheduling process of SOXS, written in Python language, using the Flask micro-framework that acts as a middle-ware between the Marshall and the ESO P2 system for the OB. 
In particular, the OBs are created through the SOXS scheduler and are stored in a standard SQL database.  In this case we opted for MySQL. The whole app is containerized using Docker and the different instances of the same container will be placed into an autoscaling group. We will make use of Amazon Elastic Container Service (ECS), with an automatic load balancer, such as Elastic Load Balancer (ELB), which will be connected to an API gateway, whose target is to intercept all the API calls and redirect them, through the ELB, to the various instances in the cluster (see High Availability section). 
For what concerns the web application we adopted the AngularJS Framework and Bootstrap toolkit for its development. However, some parts (mainly the local server) are written using JQuery and PHP.

\subsection{Interface overview}
The application opens at startup on the login form where user from specific group membership access different section of the interface. 
There are 3 groups of users defined:
\begin{itemize}
    \item\noindent\textbf{Administrator:} the main task of this group's members is the configuration of the application – i.e. the URL of the API server, and the user management (add/remove user). They also have access to the complete log of the application activity (other groups have limited access to the logs);
    \item\noindent\textbf{Science Team:} activities of this group of users is the organization of the scheduler and of the visitor execution sequence. This group has access to the log of the application activity generated by the Science team and by the operators;
    \item\noindent\textbf{Operator:} this group is active during the night and watch over how well the scheduled observations are going. It has access only to the operator's activity in the application logs.
\end{itemize}

\subsection{HTTP interface}

In order to complete its requested job, the web application needs to interface with various online services:
\begin{itemize}
    \item{The Local Server, implemented in PHP and available on the same computer, used to store and retrieve data required or produced by the application. The data are locally stored in JSON files. There are: $i$) application configuration; $ii$) user details and credential; $iii$) application activity logs.}
    \item{The SOXS Scheduler Rest API that provides a bridge to the SOXS database, the ESO P2 and the Marshall application.}
    \item{The Amazon Web Service - Simple Notification Service (AWS-SNS) that performs notifications to interested users when an urgent transient is inserted in an observation (notifications are done through various channels: sms, email, browser popup.}
\end{itemize}

An HTTP interface factory (webIntefaceFactory.js) is used in order to perform request to the API server and to the Local Server. 
This factory uses the AngularJS \$http service (based on XMLHttpRequest) to communicates with the local server and the SOXS Scheduler Rest API.
The interface with the AWS-SNS is done with the AWS Software Development Kit (SDK).

\begin{figure} [ht]
\begin{center}
\begin{tabular}{c} 
\includegraphics[height=8cm]{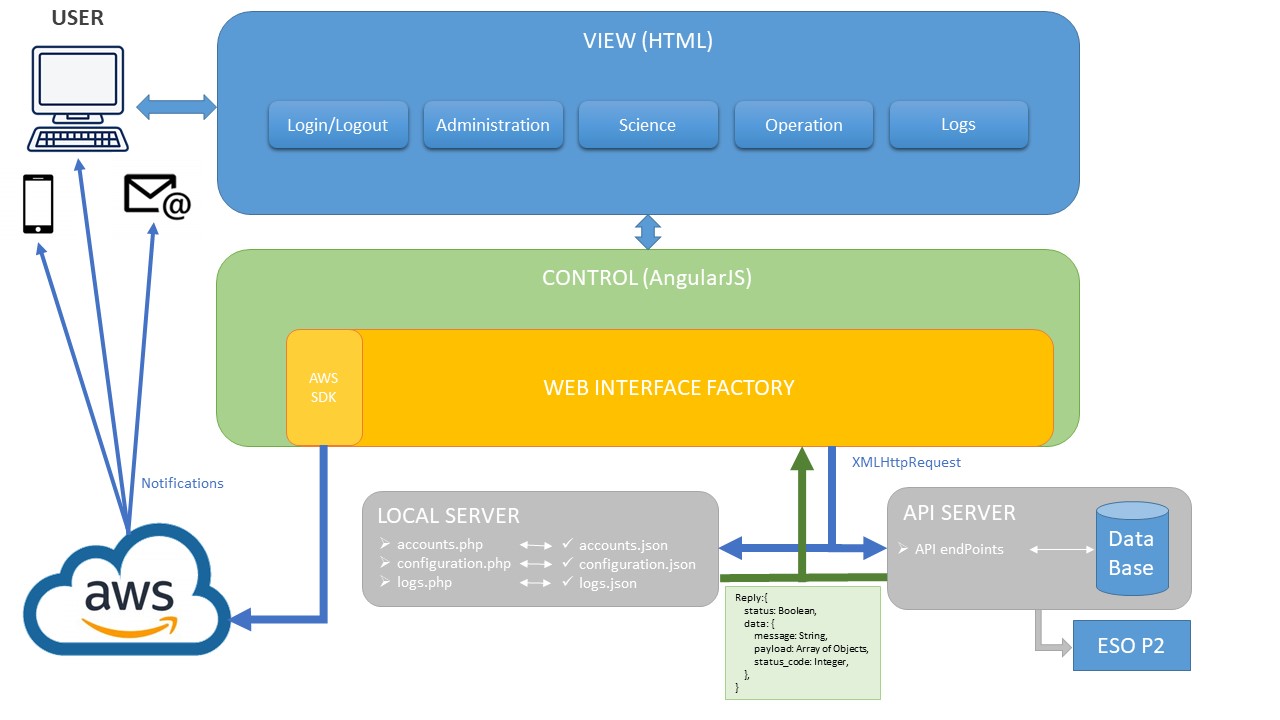}
\end{tabular}
\end{center}
\caption[httpinterface] 
{ \label{fig:httpinterface} HTTP interface.}
\end{figure}

\subsection{File and directories structure}

The file index.html includes all the stylesheets and libraries required by the application. It contains in its body a wrapper that is redirected using the angularJs routing service to the main.html file (see See Fig.~\ref{fig:pagesStructure} below).
When a logged user clicks on the “Show Logs” link, the file index.html is opened in a new tab, but the angularJs routing service redirect the wrapper to the logger.html file that has the same structure as the main.html file.
The following figure (Fig.~\ref{fig:pagesStructure}) represents the structure of the application.

   \begin{figure} [ht]
   \begin{center}
   \begin{tabular}{c} 
   \includegraphics[height=8cm]{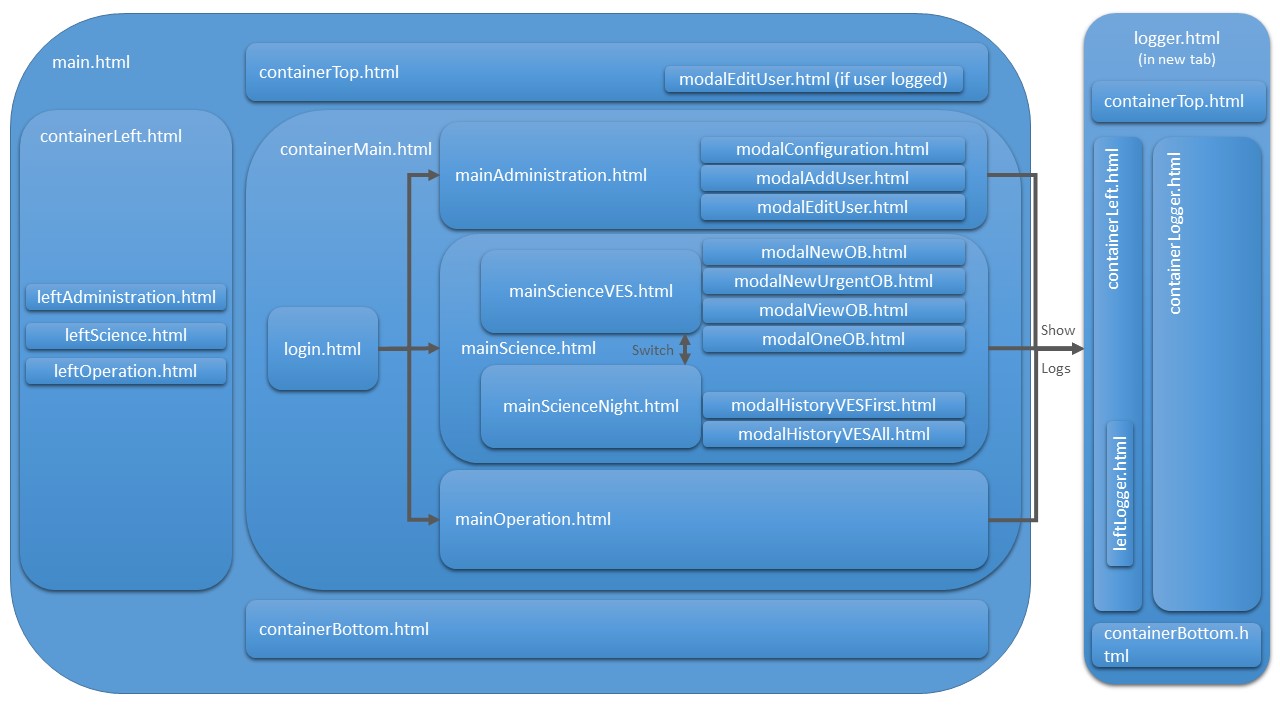}
   \end{tabular}
   \end{center}
   \caption[pagesStructure] 
   { \label{fig:pagesStructure} main.html and logger.html structure.}
   \end{figure}

\subsection{Activity Logs}

Every action occurring on the application is logged on the local server in a file in json format. This file is backed-up periodically on the API server, so only the last logs are stored locally.
A log record contains information about the action performed, the date and time of the action and the id and group of the originator.
The logs are readable by the user by clicking the “Show logs” link on the left side of the interface. This opens a new tab on the browser that displays a sortable table containing logs. The displayed logs can be filtered by date and a search facility is available to look globally for a specific string in the logs records.

   \begin{figure} [ht]
   \begin{center}
   \begin{tabular}{c} 
   \includegraphics[height=8cm]{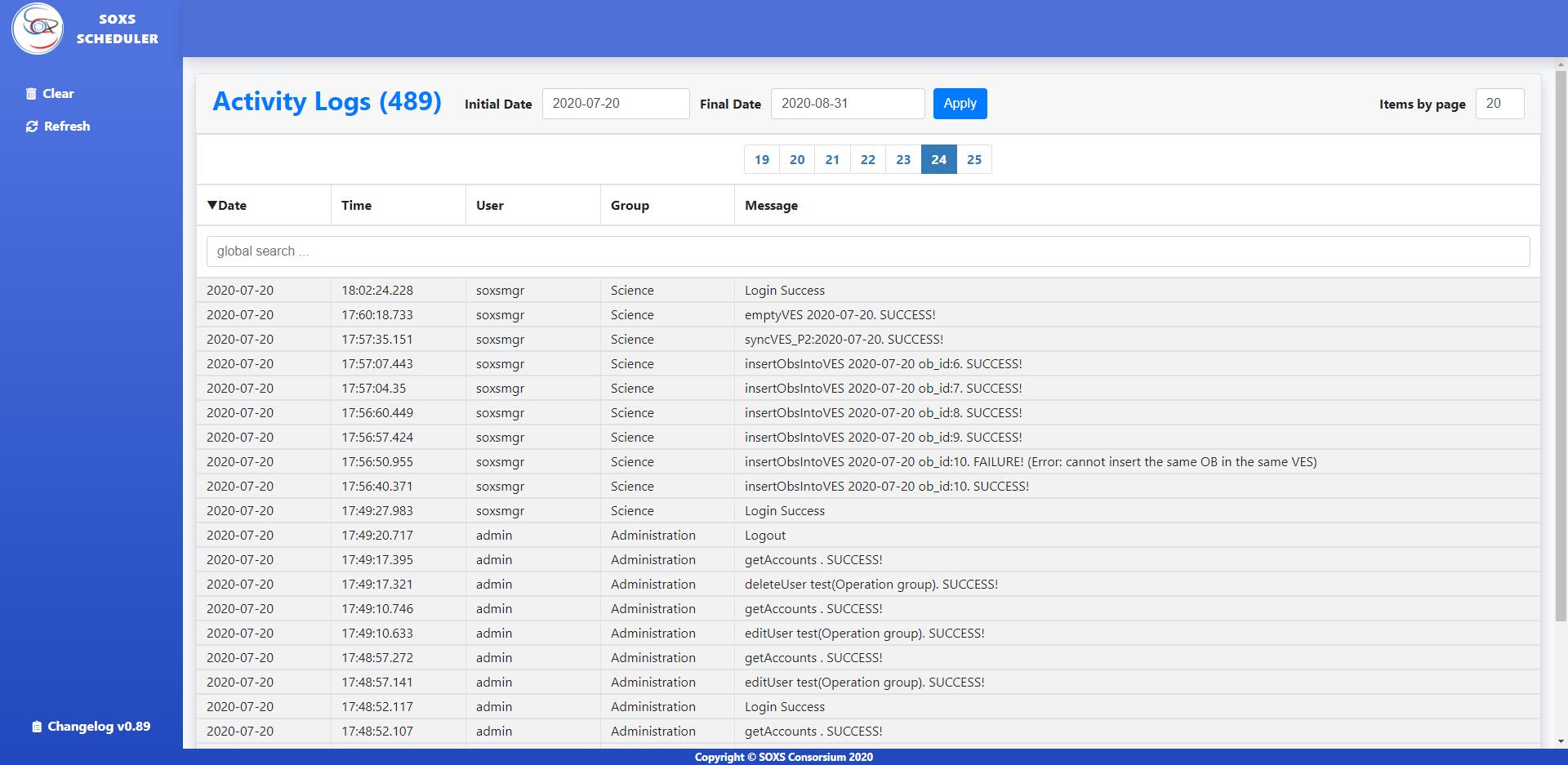}
   \end{tabular}
   \end{center}
   \caption[loggerScreen] 
   { \label{fig:loggerScreen} Logger Screen.}
   \end{figure}

\subsection{Login and accounts}

\begin{figure} [ht]
   \begin{center}
   \begin{tabular}{c} 
   \includegraphics[height=8cm]{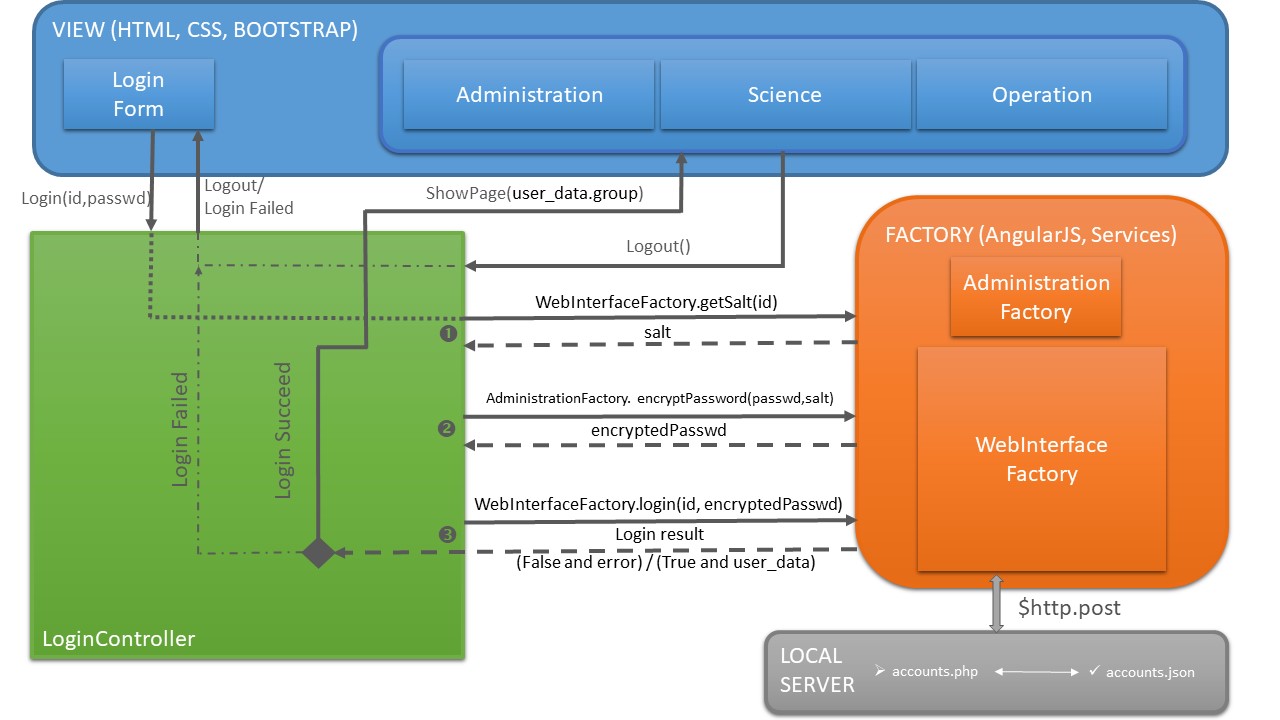}
   \end{tabular}
   \end{center}
   \caption[loginStructure] 
   { \label{fig:loginStructure} Login structure.}
   \end{figure}

The user accounts are stored in a plain json file locally. This is accessed through requests to the logs.php file on the local server by the webInterfaceFactory.
The users are divided into three groups with different roles: Administration, Science, and Operation. 
The Login form use the login controller to check the userId/password inserted and reject or accept the login, rerouting the user to the right page of their group membership.
Members of administration group are in charge of the creation of a password for every user, but each user has the possibility to change it. Password hashing uses the bcrypt function.
The logic of the login procedure is illustrated in Fig.\ref{fig:loginStructure}

\subsection{Administrator Interface}

The administration page shows a sortable, searchable table listing the users accounts registered in the system.
From this page the administrator can manage the user’s accounts:

$\bullet$   add an account;

$\bullet$   edit an existing account;

$\bullet$   delete an account;

$\bullet$   reset an account password;

$\bullet$   send email to a specific user.

From this page, an administrator has the possibility to configure some setting of the application:

$\bullet$  the Scheduler API server URL;

$\bullet$  the duration of a session before an automatic logout;

$\bullet$  the credentials used to access the AWS service used for the notifications to users.

The administrator can also view all the activity logs from all groups. The administrator has also the privilege to cancel all logs from the log file.
The following figure (Fig.\ref{fig:administrator}) shows the List of OBs page.

\begin{figure} [ht]
\begin{center}
\begin{tabular}{c} 
\includegraphics[height=8cm]{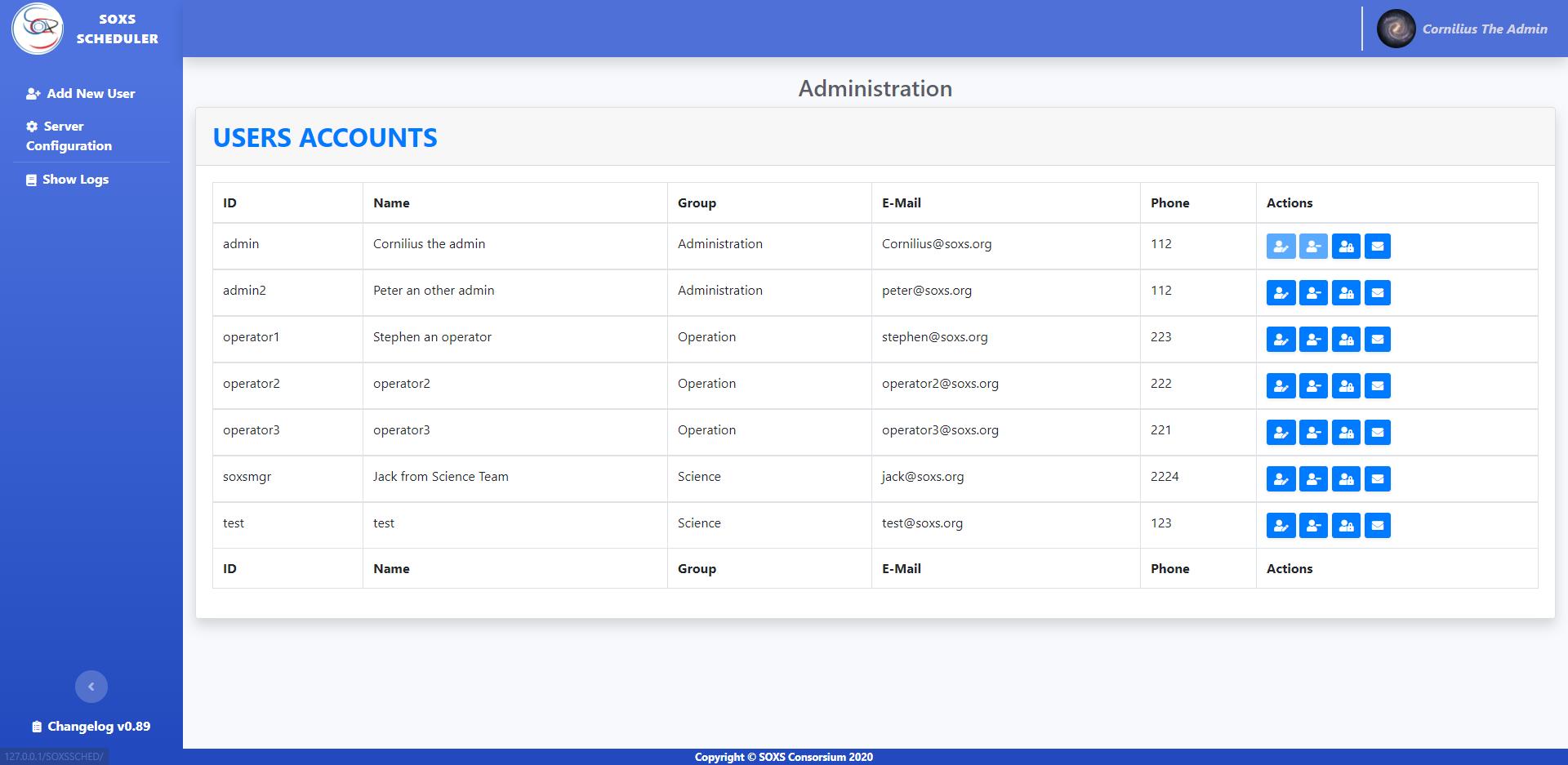}
\end{tabular}
\end{center}
\caption[administrator] 
{ \label{fig:administrator} Administration screen.}
\end{figure} 

\subsection{Science Team Interface}

When a user log in the application as member of the Science Team, he has access to the science section of the site.
At the time of writing, the science section contains 2 pages:
    
\subsubsection{List of OBs}

The list-of-OBs page displays a table with all the available observable objects present in the API server database. The table of OBs can be be sorted and filtered. 
From this page the user can performs few actions:

$\bullet$   display detailed information of a specific OB;

$\bullet$   add a new OB;

$\bullet$   add a urgent OB that should be processed as soon as possible by the operator (the operator will be informed by email, SMS and Notification on the telescope browser);

$\bullet$   remove an OB from the list;

$\bullet$   save an OB in a specific container in ESO P2;

$\bullet$   modify the magnitude of the target of an OB;

$\bullet$   synchronize the status of an OB with ESO P2;

$\bullet$   generate the finding chart of an OB;

$\bullet$   view the OB in the Marshall server;

$\bullet$   select an OB to insert it in the active Visitor Execution Sequence (VES).

The following figure (Fig.\ref{fig:listOfOBs}) shows the List of OBs page.

\begin{figure} [ht]
\begin{center}
\begin{tabular}{c} 
\includegraphics[height=8cm]{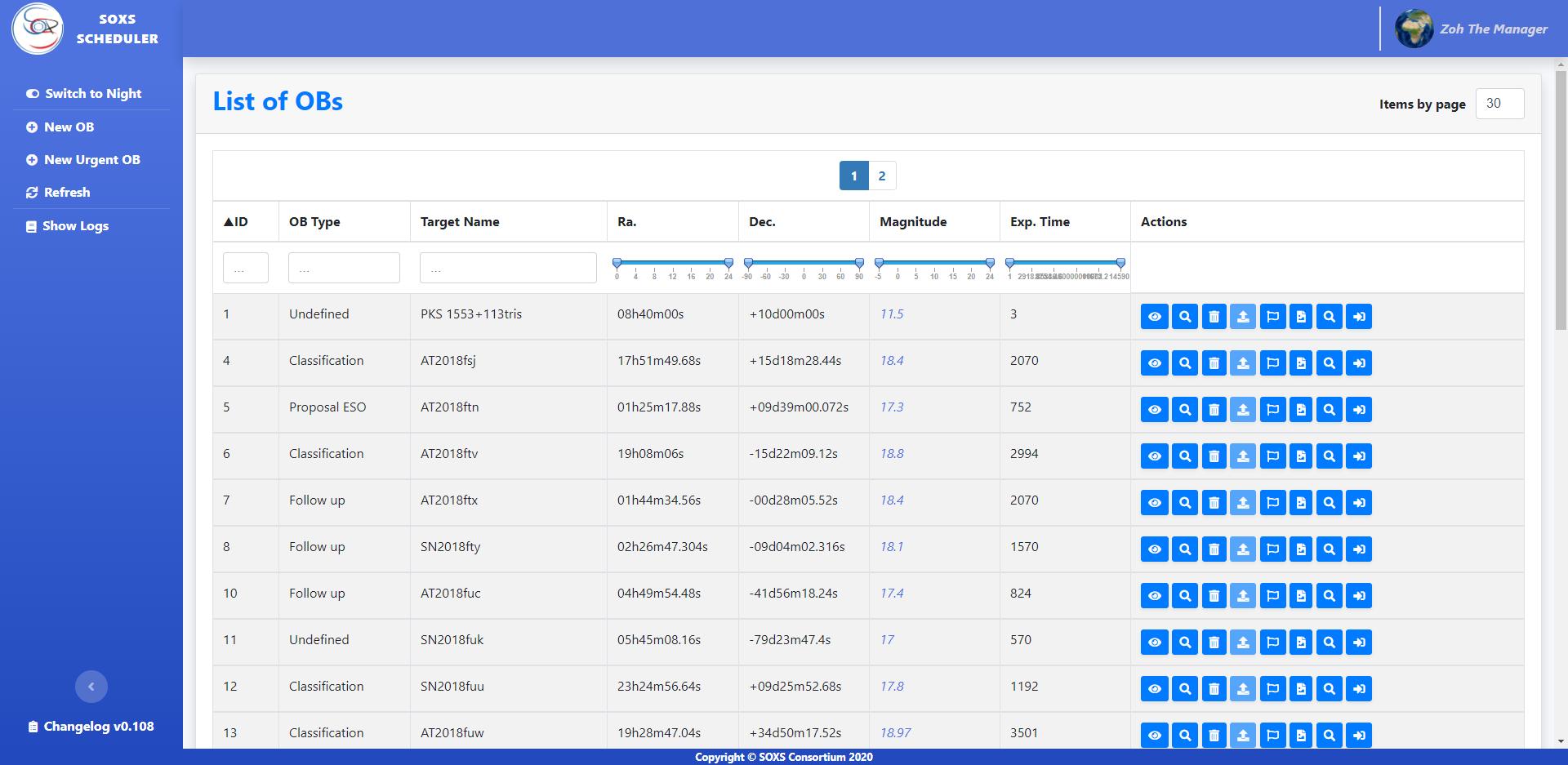}
\end{tabular}
\end{center}
\caption[listOfOBs] 
{ \label{fig:listOfOBs} List-of-OBs screen.}
\end{figure}

\subsubsection{Visitor Execution Sequence}

The VES page displays a table with all the OBs selected for the observation of a specific night. 
From this page the user can performs few actions:

$\bullet$  display detailed information of a specific OB; 

$\bullet$  modify the magnitude of the object in an OB;

$\bullet$  remove an OB from the VES;

$\bullet$  cancel the entire VES (empty it);

$\bullet$  reorder the VES (by dragging an OB to its new position);

$\bullet$  synchronize the status of an OB with ESO P2;

$\bullet$  remove an OB from the list of OBs;

$\bullet$  save an OB in a specific container in ESO P2; 

$\bullet$  generate the finding chart of an OB;

$\bullet$  view the OB in the Marshall server;

$\bullet$  synchronise the VES with ESO P2 (this operation is mandatory before the observation, a reminder is set to inform the user to do it when the VES is modified);

$\bullet$  retrieve the history of modifications performed on the VES.

The following figure (Fig.\ref{fig:ves}) shows the VES page.

   \begin{figure} [ht]
   \begin{center}
   \begin{tabular}{c} 
   \includegraphics[height=8cm]{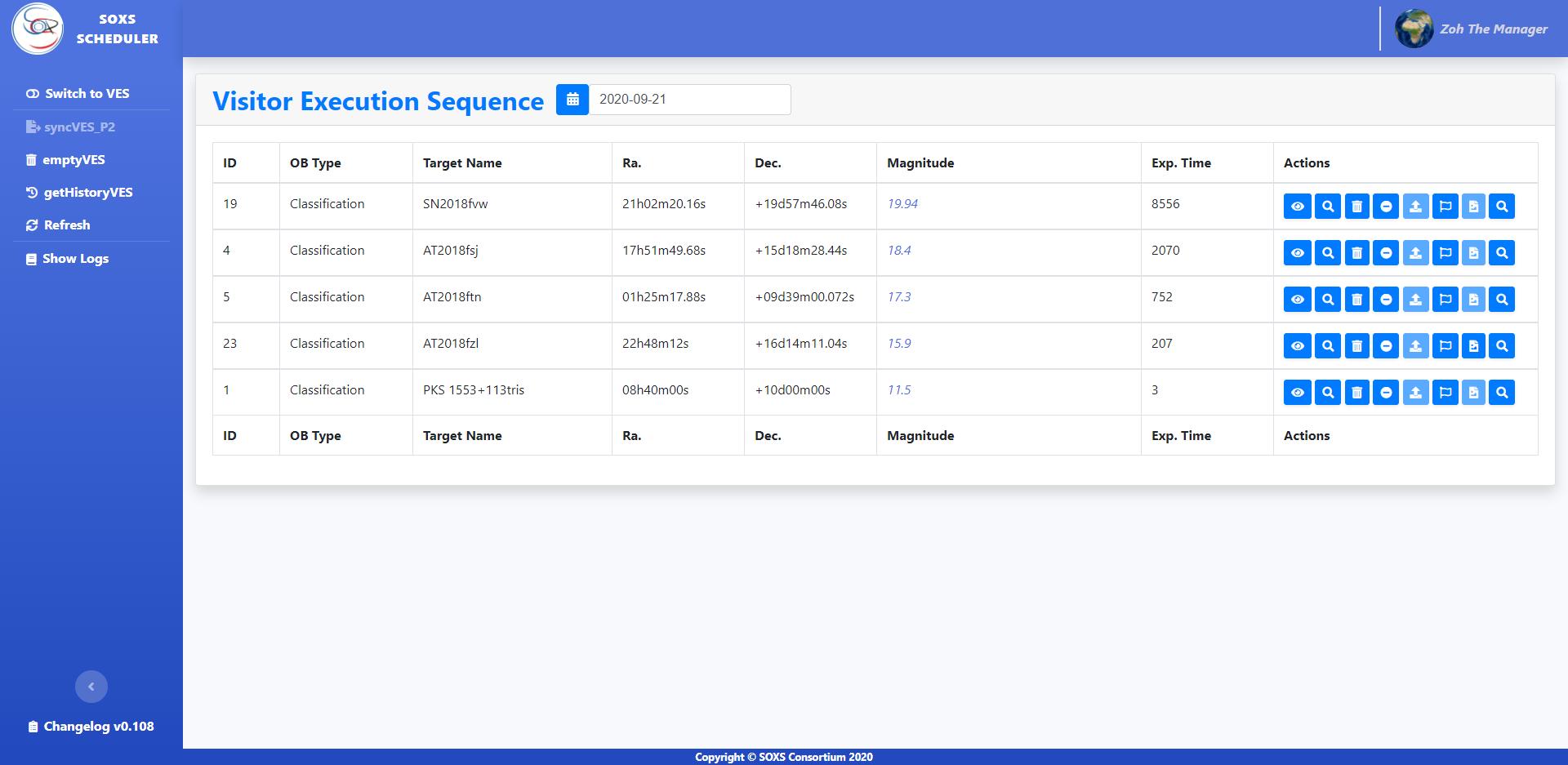}
   \end{tabular}
   \end{center}
   \caption[ves] 
   { \label{fig:ves} Visitor Execution Sequence screen.}
   \end{figure} 

\subsection{Operator Interface}

From the operator interface page the operator will execute the current VES during the night. This part of the application is not yet implemented.

\section{High availability}

Since the scheduler is in charge to quickly populate the VES (also during the night) and interface it with the surveys through the Marshall, it must be reliable and redundant, requiring a design in High Availability regime.

\begin{figure} [ht]
\centering
\includegraphics[width=0.85\textwidth]{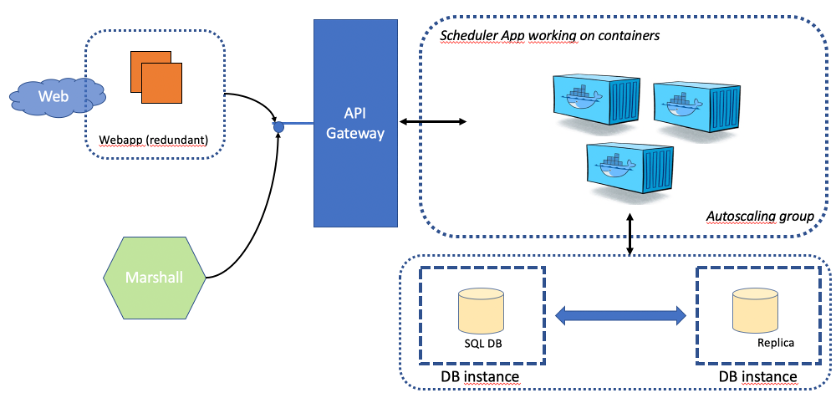}
\caption{High Availability architecture for the SOXS Scheduler.}
\label{ha}
\end{figure}

This is briefly sketched in Figure \ref{ha}. In particular, the Scheduler App is containerized (using Docker) and the different instances of the same container (used to scale up against a vast number of API calls, especially from the SOXS Marshall) are placed into an autoscaling group (a set of instances on a commercial cloud platform, like Amazon AWS) behind an API Gateway, which acts as a load balancer.
The database (an off-the-shelf relation Database management system) is replicated using a standard fail-over replica schema. This replication is intended only to support in a robust way a failure of the main Database instance.
Finally, also the Webapp (exposed by the scheduler) is intended to be replicated across more then one computer instance. The purpose of the replication is especially related to the page, seen by the operator at NTT during night time, which permits to trigger the update (or re-population) of the VES in case of observation condition changes.
We are exploring two commercial cloud platform\cite{cloud1, cloud2} vendors (Amazon Web Services and Digital Ocean) by comparing overall costs and reliability.

\bibliography{main} 
\bibliographystyle{spiebib} 

\end{document}